\newif\ifdraft
\newif\iffull
\newif\ifcomment
\newif\iflatexdiff
\newif\ifbibtex
\newif\ifpreprint
\latexdifftrue   
\preprinttrue    
\fulltrue        
\def\dvers{v1.02}
\def\snntitle{$\snn$}
\ifpreprint
\def\snntitle{$\snnbf$}
\fi
\def\dtitle{Multiplicity dependence of jet-like two-particle correlation structures in p--Pb collisions at \snntitle\ = \unit[5.02]{TeV}}
\def\stitle{Jet-like two-particle correlations in p--Pb collisions}
\ifpreprint
\documentclass[ALICE,manyauthors,12pt]{cernphprep}
\usepackage[comma,square,numbers,sort&compress]{natbib}
\usepackage{hyperref}
\else
\documentclass[final,3p,12pt]{elsarticle}
\biboptions{comma,square,numbers,sort&compress}
\usepackage{hyperref}
\fi
\usepackage{graphicx}  
\usepackage{dcolumn}   
\usepackage{bm}        
\usepackage{amssymb}   
\usepackage{amsfonts}
\usepackage{graphics}
\usepackage{grffile}   
\usepackage{epsfig}
\usepackage{units}
\usepackage[usenames]{color}
\usepackage[normalem]{ulem} 
\usepackage{color}
\usepackage[utf8]{inputenc}
\usepackage[T1]{fontenc}
\usepackage{comment}
\usepackage{enumerate}
\iflatexdiff
\RequirePackage{color}\definecolor{RED}{rgb}{1,0,0}\definecolor{BLUE}{rgb}{0,0,1}

\providecommand{\xout}[1]{{\protect\color{red}\sout{#1}}}

\fi
\ifpreprint
\else
 
\fi

\newcommand{\gevc}         {GeV/\ensuremath{c}}
\newcommand{\gevcc}         {GeV/\ensuremath{c^2}}

\newcommand{\ptt}          {\ensuremath{p_{\mathrm{T, trig}}}}
\newcommand{\pta}          {\ensuremath{p_{\mathrm{T, assoc}}}}

\newcommand{\ITS}          {\rm{ITS}}

\newcommand{\ZNA}          {\rm{ZNA}}
\newcommand{\ZNC}          {\rm{ZNC}}

\newcommand{\ZDCs}         {\rm{ZDCs}}
\newcommand{\SPD}          {\rm{SPD}}

\newcommand{\TPC}          {\rm{TPC}}
\newcommand{\VZERO}        {\rm{VZERO}}
\newcommand{\VZEROA}       {\rm{VZERO-A}}
\newcommand{\VZEROC}       {\rm{VZERO-C}}

\newcommand{\dedx}         {d$E$/d$x$}

\newcommand{\pp}           {pp}

\newcommand{\pPb}          {\mbox{p--Pb}}

\newcommand{\pt}           {\ensuremath{p_{\mathrm{T}}}{ }}

\newcommand{\snn}          {\ensuremath{\sqrt{s_{\mathrm{NN}}}}}
\newcommand{\snnbf}        {\ensuremath{\mathbf{{\sqrt{s_{\mathbf NN}}}}}}

\newcommand{\avg}[1]       {\ensuremath{\left\langle#1\right\rangle}}

\newcommand{\dd}           {\ensuremath{\mathrm{d}}}

\newcommand{\Dphi}         {\ensuremath{\Delta\varphi}}
\newcommand{\Deta}         {\ensuremath{\Delta\eta}}
\newcommand{\Ntrig}        {\ensuremath{N_{\mathrm{trig}}}}

\newcommand{\dNassoc}      {\ensuremath{\frac{\dd^2N_{\mathrm{assoc}}}{\dd\Deta\dd\Dphi}}}

\newcommand{\Eq}[1]        {Eq.~\ref{#1}}

\newcommand{\Ref}[1]       {Ref.~\citenum{#1}}

\newcommand{\red}[1]       {\textcolor{red}{#1}}

\newcommand{\warn}[1]      {{\small\textbf{\red{(!}\footnote{\textbf{\red{(!)}}~#1}\red{)}}}\marginpar{\textbf{\red{---}}}}

\newcommand{\com}[1]       {}

\newcommand{\nch}					 {{\rm N_{\rm ch}}}
\iflatexdiff

\renewcommand{\xout}[1]    {\textcolor{red}{\sout{#1}}}
 
\else

\renewcommand{\xout}[1]    {}
\fi

\graphicspath{{./img/}}
\ifdraft
\usepackage{lineno}
\linenumbers
\modulolinenumbers[5]
\usepackage{fancyhdr}
\else
\renewcommand{\warn}[1]{}

\fi
\begin{document}
\newlength{\figlen}
\setlength{\figlen}{\linewidth}
\ifpreprint
\setlength{\figlen}{0.95\linewidth}
\begin{titlepage}
\PHnumber{2014-084}                   
\PHdate{05 Jun 2014}                  
\title{\dtitle}
\ShortTitle{\stitle}
\Collaboration{ALICE Collaboration%
         \thanks{See Appendix~\ref{app:collab} for the list of collaboration members}}
\ShortAuthor{ALICE Collaboration} 
\ifdraft
\begin{center}
\today\\ \color{red}DRAFT \dvers\ \hspace{0.3cm} \$Revision: 314 $\color{white}:$\$\color{black}\vspace{0.3cm}
\end{center}
\fi
\else
\begin{frontmatter}
\title{\dtitle}
\iffull
\input{Alice_Authorlist_2014-Apr-30-PLB.tex}      
\else
\ifdraft
\author{ALICE Collaboration \\ \vspace{0.3cm}
\today\\ \color{red}DRAFT \dvers\ \hspace{0.3cm} \$Revision: 314 $\color{white}:$\$\color{black}}
\else
\author{ALICE Collaboration}
\fi
\fi
\fi
\begin{abstract}
Two-particle angular correlations between unidentified charged trigger and associated particles are measured by the ALICE detector in \pPb\ collisions at a nucleon--nucleon centre-of-mass energy of \unit[5.02]{TeV}. The transverse-momentum range 0.7 $< \pta < \ptt <$ \unit[5.0]{\gevc} is examined, to include correlations induced by jets originating from low momen\-tum-transfer scatterings (minijets).
The correlations expressed as associated yield per trigger particle are obtained in the pseudorapidity range $|\eta|<0.9$.
The near-side long-range pseudorapidity  correlations observed in high-multiplicity \pPb\ collisions  are subtracted from both near-side short-range and away-side correlations in order to remove the non-jet-like components. 
The  yields in the jet-like peaks 
 are found to be invariant  with event multiplicity with the exception of events with low multiplicity. This invariance is consistent with the particles being produced via the incoherent fragmentation of multiple parton--parton scatterings, while the yield related to the previously observed ridge structures is not jet-related.
The number of uncorrelated sources of particle production is found to increase linearly with multiplicity, suggesting no saturation of the number of multi-parton interactions even in the highest multiplicity \pPb\ collisions. 
Further, the number scales only in the intermediate multiplicity region with the number of  binary nucleon--nucleon collisions estimated with a Glauber Monte-Carlo simulation. 
\ifdraft
\ifpreprint
\end{abstract}
\end{titlepage}
\else
\end{abstract}
\end{frontmatter}
\newpage
\fi
\fi
\ifdraft
\thispagestyle{fancyplain}
\else
\end{abstract}
\ifpreprint
\end{titlepage}
\else
\end{frontmatter}
\fi
\fi
\setcounter{page}{2}


\section{Introduction}
\label{sec:intro}
Data from \pPb\ collisions at the LHC have resulted in several surprising measurements with observations  which are typically found in collisions of heavy ions and are understood to be due to a collective expansion of the hot and dense medium (hydrodynamic flow). In particular, so-called ridge structures which span over a large range in pseudorapidity ($\eta$) have been observed in two-particle correlations \cite{CMS:2012qk, alice_pa_ridge, atlasridge}.  Their modulation in azimuth is described by Fourier coefficients and  is dominated by those of second ($v_2$) and third ($v_3$) order \cite{alice_pa_ridge, atlasridge, Chatrchyan:2013nka}. They are also found in the correlations of four particles \cite{Chatrchyan:2013nka, Aad:2013fja} which are less sensitive to non-flow effects like resonance decays and jets.
Evidence for the existence of a common flow velocity field has been further corroborated by particle-identification measurements of the same observables \cite{ABELEV:2013wsa}. They revealed that the $v_2$ of pions, kaons and protons as a function of $\pt$ shows a characteristic mass ordering as well as a crossing of pion and proton $v_2$ at about \unit[2.5]{\gevc} which is reminiscent of measurements in Pb--Pb collisions \cite{Abelev:2012di}.
These findings hint at potentially novel mechanisms in collisions of small systems which are far from being understood theoretically. Several authors describe the results in the context of hydrodynamics \cite{Bozek:2011if, Bozek:2012gr, Bozek:2013uha, Werner:2013ipa, Werner:2013tya}, but also explanations in the framework of saturation models successfully describe some of the measurements \cite{Dusling:2012wy, Dusling:2013oia}. 

While  measurements of these correlations are suggestive of similarities between Pb--Pb and \pPb\ collisions, measurements sensitive to energy loss in a hot and dense medium reveal no or minor modifications with respect to pp collisions. The inclusive hadron nuclear modification factor $R_{\rm pA}$ of minimum-bias \pPb\ events shows no significant deviations from unity up to \unit[20]{\gevc} \cite{ALICE:2012mj}. Measurements of the dijet transverse momentum imbalance show comparable results to simulated pp collisions at the same center-of-mass energy, independent of the forward transverse energy \cite{Chatrchyan:2014hqa}. 

Towards a more complete picture of the physical phenomena involved in \pPb\ collisions, it is interesting to study QCD interactions in the $\pt$ range where these  ridge-like structures have been observed. 
Parton scatterings with large transverse-momentum transfer ($Q^2 \gg \Lambda_{\rm QCD}$, typically called {\it hard interactions)}
lead to phenomena such as high-$\pt$ jets.
QCD-inspired models extrapolate these
interactions to the low-$\pt$ region where several such interactions can occur per nucleon--nucleon collision (multiple parton interactions -- MPIs) and can hence contribute significantly 
to particle production \cite{PhysRevD.84.034026, JF}. 
The objective of the analysis presented in this paper is to investigate if jet-like structures in this low-\pt region show modifications as a function of event multiplicity in addition to the appearance of the ridge-like structures.  The analysis employs two-particle azimuthal correlations within $|\eta|<0.9$ from low to intermediate transverse momentum ($0.7 < \pt <$~\unit[5]{\gevc}) in \pPb\ collisions. After subtraction of the long-range pseudorapidity ridge-like structures, the yields of the jet-like near- and away-side peaks are studied as a function of multiplicity. 
As already shown in pp collisions, this analysis procedure allows the extraction of the so-called number of uncorrelated seeds, which in PYTHIA is proportional to the number of MPIs \cite{Abelev:2013sqa}. Thus the presented results allow to draw conclusions on the contribution 
of hard processes to particle production as a function of event multiplicity.

The paper is structured as follows: Section~\ref{sec:setup} presents the experimental setup followed by the event and track selections in Section~\ref{sec:selection} and the analysis details in Section~\ref{sec:analysis}. The results are presented in Section~\ref{sec:results} followed by a summary.

\section{Experimental setup}
\label{sec:setup}
In the present analysis, \pPb\ collision data at a centre-of-mass energy of $\snn=$~\unit[5.02]{TeV} collected by the ALICE detector in 2013 are used.
The energies of the beams were \unit[4]{TeV} for the proton beam and \unit[1.58]{TeV} per nucleon for the lead beam. The nucleon--nucleon centre-of-mass system moves with respect to the ALICE laboratory system with a rapidity of $-0.465$, i.e.\ in the direction of the proton beam. In the following, $\eta$ denotes the pseudorapidity in the laboratory system.

A detailed description of the ALICE detector can be found in \Ref{Aamodt:2008zz}. 
The  subdetectors used in the present analysis for charged particle tracking are the Inner Tracking System~(\ITS) and the Time Projection Chamber~(\TPC), both operating in a solenoidal magnetic field of \unit[0.5]{T} and covering a common acceptance of $|\eta| < 0.9$. 
The \ITS\ consists of six layers of silicon detectors: two layers of Silicon Pixels Detectors~(\SPD), two layers of Silicon Drift Detectors and two layers of Silicon Strip Detectors, from the innermost to the outermost ones. 
The \TPC\ provides tracking and particle identification by measuring the curvature of the tracks in the magnetic field and the specific energy loss \dedx. 
The VZERO detector, which consists of two arrays of 32 scintillator tiles each, covers the full azimuth within $2.8 < \eta < 5.1$ (\VZEROA) and $-3.7 < \eta < -1.7$ (\VZEROC) and is used for triggering, event selection and event characterization. The trigger requires a signal of logical coincidence in both \VZEROA\ and \VZEROC. The \VZEROA, located in the flight direction of the Pb ions, is used to define event classes corresponding to different particle-multiplicity ranges.
In addition, two neutron Zero Degree Calorimeters~(\ZDCs), located at \unit[112.5]{m}~(\ZNA) and \unit[$-112.5$]{m}~(\ZNC) from the interaction point, are used for the event selection. 
The ZNA has an acceptance of 96\% for neutrons originating from the Pb nucleus and the deposited energy is used as an alternative approach to define the event-multiplicity classes.

\section{Event and track selection}
\label{sec:selection}
The employed event selection \cite{ALICE:2012xs} accepts 99.2\% of all non-single-diffractive collisions.
Beam-induced background is removed by a selection on the signal amplitude and arrival times in the two \VZERO\ detectors.
The primary vertex position is determined from the tracks reconstructed in the \ITS\ and \TPC\ as described
in \Ref{Abelev:2012hxa}. The vertex reconstruction algorithm is fully efficient for events with at least
one reconstructed primary charged particle in the common \TPC\ and \ITS\ acceptance.
Events with the coordinate of the reconstructed vertex along the beam axis $z_{\rm vtx}$ within \unit[10]{cm} from the nominal interaction point are selected.
About $8 \cdot 10^7$ events pass these event selection criteria and are used for the analysis.

The analysis uses charged-particle tracks reconstructed in the \ITS\ and \TPC\ with $0.2<\pt<$ ~\unit[5]{\gevc} within a fiducial region of $|\eta|<\eta_{\rm max}$ with $\eta_{\rm max} = 0.9$. 
The track selection is the same as in \Ref{alice_pa_ridge} and is based on selections on the number of space points, the quality of the track fit and the number of hits in the \ITS, as well as the Distance of Closest Approach (DCA) to the reconstructed collision vertex. The track selection is varied in the analysis for the study of systematic uncertainties~\cite{alice_pa_ridge}.

The efficiency and purity of the track reconstruction and the track selection
for primary charged particles (defined as the prompt particles
produced in the collision, including decay products, except those from
weak decays of strange particles) are estimated from a Monte-Carlo
simulation using the DPMJET version 3.05 event generator~\cite{Roesler:2000he} with particle transport
through the detector using GEANT3~\cite{geant3ref2} version 3.21. The efficiency and acceptance for track reconstruction is 68--80\% for the $\pt$ range \unit[0.2--1]{\gevc}, and 80\% for $\pt >$ \unit[1]{\gevc} with the aforementioned track selections. The reconstruction performance is independent of the \pPb\ event multiplicity.
The remaining contamination from secondary particles due to interactions in the detector material and weak decays decreases from about 5\% to 1\% in the $\pt$ range from 0.5 to \unit[5]{\gevc}.
The contribution from fake tracks, false associations of detector signals, is negligible.
Corrections for these effects are discussed in Section~\ref{sec:analysis}.
Alternatively, efficiencies are estimated using HIJING version 1.36 \cite{hijing} with negligible differences in the results.

In order to study the multiplicity dependence of the two-particle correlations, the events are divided into classes defined according to the charge deposition in the \VZEROA\ detector (called V0A when referring to it as a multiplicity estimator). The events are classified in  5\% percentile ranges of the multiplicity distribution, denoted as ``0--5\%'' to ``95--100\%'' from the highest to the lowest multiplicity.

The \VZEROA\ detector is located in the direction of the Pb beam and thus sensitive to the fragmentation of the Pb nucleus, and is used as default multiplicity estimator. Two other estimators are employed to study the behaviour of the two-particle correlations as a function of the $\eta$-gap between the detector used to measure the multiplicity and the tracking detectors. These are CL1, where the signal is taken from the outer layer of the \SPD\ ($|\eta|$ < 1.4), and ZNA, which uses the \ZNA\ detector ($|\eta| > 8.8$).
Due to the limited efficiency of the ZNA, results are only presented
for the 95\% highest-multiplicity events.
These estimators select events with different ranges of multiplicity at midrapidity. While the V0A estimator selects event classes with on average about 5 to 69 charged particles within $|\eta|<0.9$ and $\pt$ larger than \unit[0.2]{\gevc}, the CL1 has a slightly larger range (about 2 to 78) and the ZNA has a smaller range (about 10 to 46). 

The observables in this analysis are calculated for events with at least one particle with \linebreak $\pt>$~\unit[0.2]{\gevc} within $|\eta|<0.9$. Monte-Carlo simulations show that this selection reduces the number of events compared to all inelastic events by about 2\%. These events are concentrated at low multiplicity in the 80--100\% multiplicity classes.

\section{Analysis}
\label{sec:analysis}

The two-particle correlations between pairs of trigger and associated charged particles are expressed as the associated yield per trigger particle in a given interval of transverse momentum, for each multiplicity class. The associated per-trigger yield is measured as a function of the azimuthal difference $\Dphi$ (defined within $-\pi/2$ and $3\pi/2$) and of the pseudorapidity difference $\Deta$. 
The condition $\pta < \ptt$ between transverse momenta of trigger and associated particles  is required.

The associated yield per trigger particle is defined as
\begin{equation}
\frac{1}{\Ntrig} \dNassoc = S(\Deta,\Dphi) \cdot C(\Deta,\Dphi), \label{pertriggeryield} 
\end{equation}
where $\Ntrig$ is the total number of trigger particles in the event class and $\pt$ interval.
The signal distribution
$S(\Deta,\Dphi) = 1/\Ntrig\ \dd^2N_{\rm same}/\dd\Deta\dd\Dphi$
is the associated yield per trigger particle for particle pairs from the same event.
The correction factor $C$ is defined as:
\begin{equation}
C(\Deta,\Dphi) = \frac{\widetilde{B}(\Deta)}{B(\Deta,\Dphi)},
\end{equation}
where $B$ describes the pair acceptance and pair efficiency of the detector while $\widetilde{B}$ is the pair acceptance of a perfect but pseudorapidity-limited detector, i.e. a triangular shape defined by $\widetilde{B}(\Deta) = 1 - |\Deta|/(2 \cdot \eta_{\rm max})$.
In this way, the resulting associated yields per trigger particle count only the particles entering the detector acceptance, as it is required for the definition of uncorrelated seeds, see below and the detailed discussion in \Ref{Abelev:2013sqa}.

$B(\Deta,\Dphi) = \alpha\ \dd^2N_{\rm mixed}/\dd\Deta\dd\Dphi$ is constructed by correlating the trigger particles in one event with the associated particles from
different events in the same  multiplicity class and within the same \unit[2]{cm}-wide $z_{\rm vtx}$ interval
(each event is mixed with about 5--20 events).
It is normalized with a factor $\alpha$ which is chosen such that $B(\Deta,\Dphi)$ is unity at $\Dphi = \Deta\approx 0$ for pairs where both particles travel in approximately the same direction.
The yield defined by \Eq{pertriggeryield} is constructed for each $z_{\rm vtx}$ interval to account for differences in pair acceptance and in pair efficiency.
After efficiency correction (described below) the final per-trigger yield is obtained by calculating the average of the $z_{\rm vtx}$ intervals weighted by $\Ntrig$.
A selection on the opening angle of the particle pairs is applied in order to avoid a bias due to the reduced efficiency for pairs with small opening angles. Pairs are required to have a separation of $|\Delta\varphi^{*}_{\rm min}|>$~\unit[0.02]{rad} or $|\Deta|>0.02$, where $\Delta\varphi^*_{\rm min}$  is
the minimal azimuthal distance at the same radius between the two tracks within the active
detector volume after accounting for the bending in the magnetic field.

Furthermore, correlations induced by secondary particles from neutral-particle decays are suppressed by cutting on the invariant mass ($m_{\rm inv}$) of the particle pair. In this way pairs are removed which are likely to stem from a $\gamma$-conversion ($m_{\rm inv} <$~\unit[0.04]{\gevcc}), a K$^0_s$ decay ($|m_{\rm inv} - m({\rm K}^0)| <$~\unit[0.02]{\gevcc}) or a
$\Lambda$ decay ($|m_{\rm inv} - m(\Lambda)| <$~\unit[0.02]{\gevcc}).
The corresponding masses of the decay particles (electron, pion, or pion/proton) are assumed in the $m_{\rm inv}$ calculation.

Each trigger and each associated particle is weighted with a correction factor that accounts for reconstruction efficiency and contamination by secondary particles. These corrections are applied as a function of $\eta$, $\pt$ and $z_{\rm vtx}$. 
The correction procedure is validated by applying it to simulated events and comparing the per-trigger pair yields with the input Monte-Carlo simulations. 
The remaining difference after all corrections (Monte-Carlo non-closure) is found to be negligible.

\subsection{Long-Range Correlations Subtraction}

In addition to the jet-like peaks, ridge structures have  been observed in \pPb\ collisions \cite{alice_pa_ridge}\cite{atlasridge}. These long-range structures are mostly independent of $\Deta$ outside the jet-like peak and assumed to be independent below the peak and their modulation in azimuth is described by a Fourier expansion up to the third order. To study the properties of the jet-like peaks, these structures are subtracted.

\begin{figure}[ht!f]
\centering
\includegraphics[width=\textwidth]{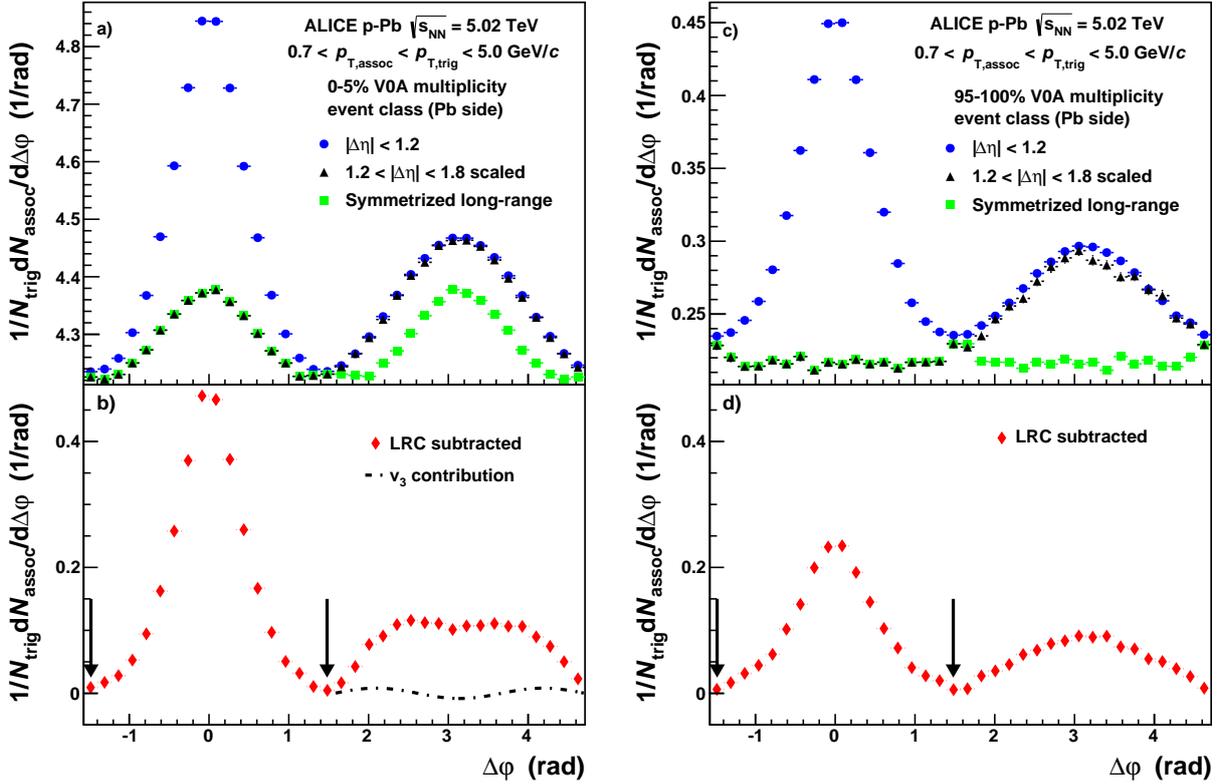}
\caption{\label{fig:pertriggeryields}
Per-trigger yield as a function of $\Dphi$ with $0.7<\pta < \ptt <$~\unit[5]{\gevc} in the 0--5\% event class (left) and 95--100\% event class (right). The distributions show the correlations before subtraction (blue circles), the long-range correlations (black triangles) scaled according to the $\Deta$ region in which they are integrated, the symmetrized near-side long-range correlations (green squares) and the correlations after long-range correlations (LRC) subtraction (red diamonds). The vertical arrows indicate the integration regions while the curve in the bottom left panel shows the magnitude of the third Fourier component on the away side. Statistical uncertainties are shown but are smaller than the symbol size.
}
\end{figure}

On the near side ($-\pi/2 < \Dphi < \pi/2$), the jet-like peak is centered around $(\Deta = 0, \Dphi = 0)$, while the ridge structures extend to large $\Deta$. Thus the near side is divided into short-range ($|\Deta| < 1.2$) and long-range ($1.2 < |\Deta| < 1.8$) correlations regions which are correctly normalized and subtracted from one another. Figure~\ref{fig:pertriggeryields} shows the $\Dphi$-distributions of the per-trigger yield in these two regions in the highest (0--5\%) and lowest (95--100\%) multiplicity classes. 

On the away side ($\pi/2 < \Dphi < 3\pi/2$) the  jet contribution is also elongated in $\Deta$.  The jet and ridge contribution can therefore not be disentangled.  
As the ridge structures are  mostly symmetric around $\Dphi = \pi/2$ (the second Fourier coefficient is four times larger than the third coefficient \cite{alice_pa_ridge, atlasridge}), the near-side long-range correlations are mirrored around $\Dphi = \pi/2$ and  subtracted from the away side (measured in $|\Deta| < 1.8$). Also shown in Fig.~\ref{fig:pertriggeryields} are the $\Dphi$-distributions of the symmetrized long-range correlations and the correlations after subtraction. 
Obviously, this symmetrization procedure does not account correctly for odd Fourier coefficients. To assess the effect of the third coefficient on the extracted observables, an additional $2v_3^{2} \cos 3\Dphi$ functional form is subtracted before the symmetrization.
The $v_3$ is estimated as a function of multiplicity with the subtraction procedure described in \Ref{alice_pa_ridge}. 
The influence of the $v_3$ contribution is illustrated in the bottom left panel of Fig.~\ref{fig:pertriggeryields}.
The effect of the symmetrization of the third Fourier component on the away-side yield  amounts up to 4\% and is a major contribution to the systematic uncertainties.

\subsection{Observables}

The event-averaged near-side, $\langle N_{\rm assoc,near side} \rangle$, and away-side, $\langle N_{\rm assoc,away side} \rangle$, per-trigger yields are sensitive to the fragmentation properties of low-\pt partons. They are calculated as the integral of the $\Dphi$ projection of the long-range subtracted per-trigger yield (bin counting) respectively in the near-side and away-side peaks, above the combinatorial background.
By definition after subtracting the long-range correlations ($1.2 < |\Deta| < 1.8$) from the short-range one ($|\Deta| < 1.2$), the baseline should be zero. 
Nevertheless, owing to minor differences between the detector efficiencies and those estimated with the Monte-Carlo simulations and a slight dependence of the single-particle distribution on $\eta$, a small residual baseline is present (about 0.003, hardly visible in Fig.~\ref{fig:pertriggeryields}), which is taken into account.
Fig.~\ref{fig:pertriggeryields} shows that the away-side peak is slightly wider than the near-side peak. Therefore, the near-side yield is evaluated in the region $|\Dphi| < 1.48$ and the away-side yield in $|\Dphi| > 1.48$. 
For the systematic uncertainty estimation, the value $1.48$ has been varied by $\pm 0.09$.

Alternatively, the yields are also calculated with a fit method, using  two Gaussians  on the near side and one Gaussian on the away side superimposed on a constant baseline \cite{Abelev:2013sqa}. The differences between the results obtained with the two methods are included in the systematic uncertainties.

The average number of trigger particles depends on the number of parton scatterings per event as well as on the fragmentation properties of the partons. 
Therefore, the ratio between the number of trigger particles and the per-trigger yields is computed
with the goal to reduce the dependence on fragmentation properties.
This ratio, called average number of uncorrelated seeds, is defined for symmetric \pt bins as:
\begin{equation}
\avg{N_{\rm uncorrelated\ seeds}} = \frac{\avg{N_{\rm trig}}}{\avg{N_{\rm correlated\ triggers}}} = \frac{\avg{N_{\rm trig}}}{1+\avg{N_{\rm assoc,near side}}+\avg{N_{\rm assoc,away side}}}, \label{eq_us}
\end{equation}
where the correlated triggers are calculated as the sum of the trigger particle  and the  particles  associated to that trigger particle. In PYTHIA, for \pp\ collisions~\cite{Abelev:2013sqa}, the uncorrelated seeds are found to be linearly correlated to the number of MPIs in a certain \pt range, independent of the $\eta$ range explored. The selection $\pt >$~\unit[0.7]{\gevc} has been found  optimal since it is close to $\Lambda_{\rm QCD}$ and high enough to reduce contributions of hadrons at  low \pt, e.g. from resonances and string decays.

\subsection{Systematic Uncertainties}

Table~\ref{tab:systematics} summarizes the systematic uncertainties related to the near-side and away-side long-range-subtracted yields extraction and to the uncorrelated seeds calculation. 
The largest uncertainty (5\%) for the yields is due to the integration method estimated from the difference between bin counting and the fit. The $v_3$-component estimation gives rise to an uncertainty only on the away side which is multiplicity-dependent. It is indicated by the range in the table where the largest value of 4\% is obtained for the highest multiplicity. 
Other non-negligible uncertainties are due to the track selection (2\%), the pile-up contamination (1\%), estimated by excluding the tracks from different colliding bunch crossings, and the uncertainty on the tracking efficiency (3\%) \cite{ALICE:2012mj}.

The total uncertainty for the yields is 6--8\%, which translates into 3\% uncertainty for the uncorrelated seeds where, owing to the definition, some uncertainties cancel. The total uncertainty is mostly correlated between points and between the different estimators.

\begin{table}[bht!f] \centering
  \caption{\label{tab:systematics}
    Summary of the systematic uncertainties. The uncertainties are independent of multiplicity, apart from the effect of the third Fourier component $v_3$.
    }
  \begin{tabular}{cccc}
    \hline
    Source        & Near-side yield   & Away-side yield & Uncorrelated seeds\\
    \hline
      Bin counting vs. fit  			&     5\% &   5\%& 1\%\\
      Baseline estimation  			&      negl.&  1\%& negl.\\
      $v_3$ component  				&      0\% &  0--4\% & 0--1\%\\
      Track selection				&      2\%&   2\%&negl.\\
      Tracking efficiency  		  	&      3\%&   3\%&3\%\\
      Pile-up  					&      1\%&  1\% &negl.\\
      MC closure  				&      negl. & negl. & negl.\\
      Event generator  				&      negl. &   negl.& negl.\\
    \hline
     Total						&      6\% & 6--8\% & 3\%\\
  \end{tabular}
\end{table}

\section{Results}
\label{sec:results}

\begin{figure}[ht!f]
\centering
\includegraphics[width=\textwidth]{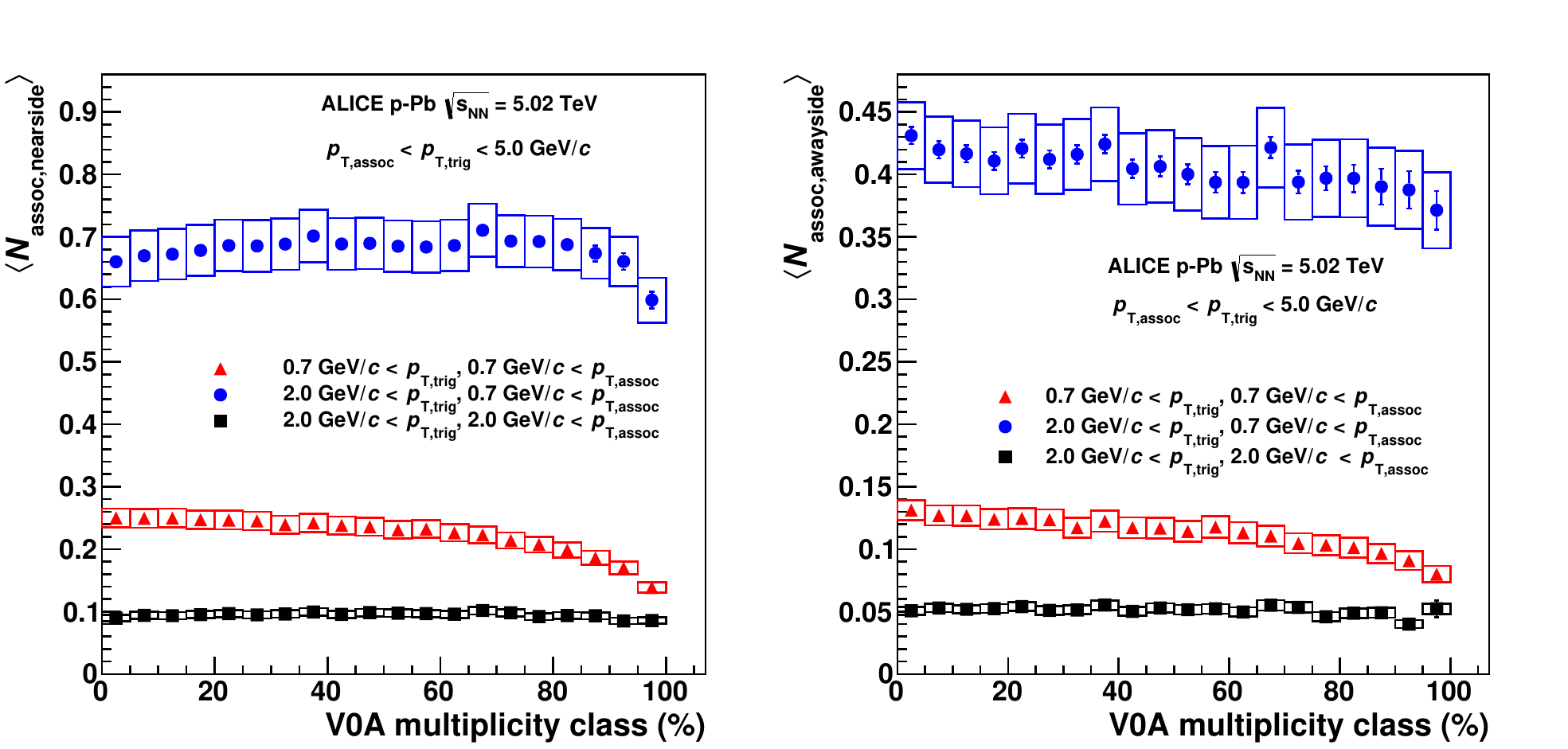}
\caption{\label{fig:near_away_V0A}
Near-side (left panel) and away-side (right panel) per-trigger yields after long-range correlations subtraction as a function of V0A multiplicity class for several \pt cuts for trigger and associated particles: \unit[0.7--5.0]{\gevc} (red triangles), \unit[0.7--5.0]{\gevc} for $\pta$ and \unit[2--5]{\gevc} for $\ptt$ (blue circles) as well as \unit[2--5]{\gevc} (black circles). Statistical (lines) and systematic uncertainties (boxes) are shown, even though the statistical ones are mostly smaller than the symbol size.
}
\end{figure}

The near-side and away-side per-trigger yields are shown in Fig.~\ref{fig:near_away_V0A} as a function of V0A multiplicity class for three different \pt ranges. For the range \unit[0.7]{\gevc} $< \pta  < \ptt <$ \unit[5.0]{\gevc} (red triangles), the near-side (away-side) per-trigger yield increases from about 0.14 (0.08) in the lowest multiplicity class up to about 0.25 (0.12) at 60$\%$, and it remains nearly constant from 60$\%$ to the highest multiplicity class. 

The trigger particles can originate both from soft and hard processes, while the associated particles mostly belong to the minijets which originate from hard processes. Therefore, in the region where the associated yields per trigger particle show a plateau, the hard processes and the number of soft particles must exhibit the same evolution with multiplicity.
This can be more easily understood with an example event containing $N_{\rm minijets}$ with $N_{\rm assoc}$ associated particles each and a background of $N_{\rm soft}$ particles with no azimuthal correlation. In this scenario, the associated yield per trigger-particle is:
\begin{equation}
\frac{\mbox{associated yield}}{\mbox{trigger particle}} = \frac{N_{\rm minijets} \cdot N_{\rm assoc}(N_{\rm assoc}-1)/2}{N_{\rm minijets} \cdot N_{\rm assoc} + N_{\rm soft}}.
\end{equation}
When the overall multiplicity, i.e. the denominator, changes, the fraction is constant if $N_{\rm minijets}$ (hard processes) and $N_{\rm soft}$ (soft processes) increase by the same factor. The given example can be easily extended to several events and to a different number of associated particles per minijet.

Increasing the \pt threshold of the trigger particles to \unit[2]{\gevc} (blue circles in Fig.~\ref{fig:near_away_V0A}), results in larger yields but with qualitatively the same multiplicity dependence. The plateau region extends in this case up to the 80\% multiplicity class. Increasing also the threshold for the associated particles to \unit[2]{\gevc} (black squares) reduces the yields while the plateau  remains over a wide multiplicity range.

\begin{figure}[ht!f]
\centering
\includegraphics[width=\textwidth]{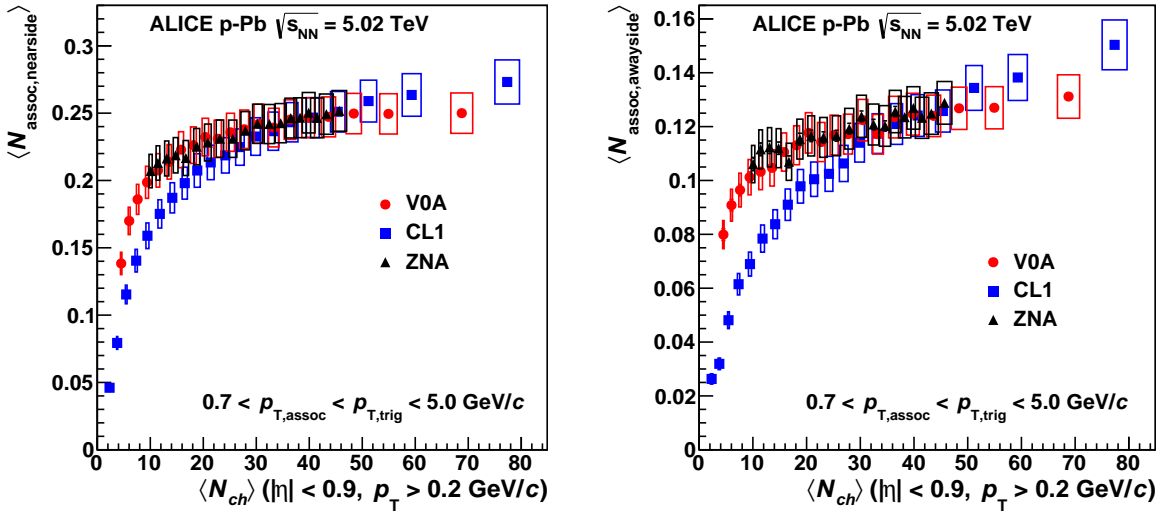}
\caption{\label{fig:near_away_estimators}
Near-side (left panel) and away-side (right panel) per-trigger yields after long-range correlations subtraction as a function of the midrapidity charged particle multiplicity for the V0A (red circles), CL1 (blue squares) and ZNA (black triangles) multiplicity estimators. Statistical (lines) and systematic uncertainties (boxes) are shown, even though the statistical ones are smaller than the symbol size.
}
\end{figure}

To compare results obtained with different multiplicity estimators, for each multiplicity class the average number of charged particles at midrapidity ($|\eta|<0.9$) with $\pt >$ \unit[0.2]{\gevc} has been computed. 
Figure~\ref{fig:near_away_estimators} shows the per-trigger yields in the near-side and in the away-side peaks as a function of the midrapidity charged particle multiplicity for the standard estimator V0A as well as for CL1 and ZNA. 
As discussed above, the multiplicity range covered by these estimators depends on the separation in pseudorapidity of the estimator and the tracking detector. The near-side (away-side) yields for V0A and ZNA show the same behaviour in the region between 10 and 45 charged particles in which  their multiplicity range overlaps: a mild increase from about 0.2 (0.1) to about 0.25 (0.13). 
Below 10 charged particles, the yields for V0A decrease significantly to about 0.14 on the near side and 0.08 on the away side.
The yields for CL1 exhibit a steeper slope than the two other estimators. This behaviour is expected from the event-selection bias imposed by the overlapping $\eta$-region of event selection and tracking: on the near side (away side) the value increases from about 0.04 to 0.27 (from about 0.02 to 0.15).
The CL1 trends are qualitatively consistent with the results in \pp\ collisions~\cite{Abelev:2013sqa}.
The overall behaviour for each estimator is similar when using higher \pt cuts for associated and trigger particles.

A key step of the analysis procedure is the subtraction of the long-range correlations. To assess the effect of this subtraction, a comparison between the yields with and without the  ridge contribution has been performed.
The determination of the yields in these two cases is, however, slightly different, since the non-subtracted distribution does not have a zero baseline by construction. 
In this case, the baseline is determined in the long-range correlations region ($1.2 < |\Deta| < 1.8$)  between the near-side ridge and the away-side peak at $1.05 < |\Dphi| < 1.22$.

\begin{figure}[ht!f]
\centering
\includegraphics[width=\textwidth]{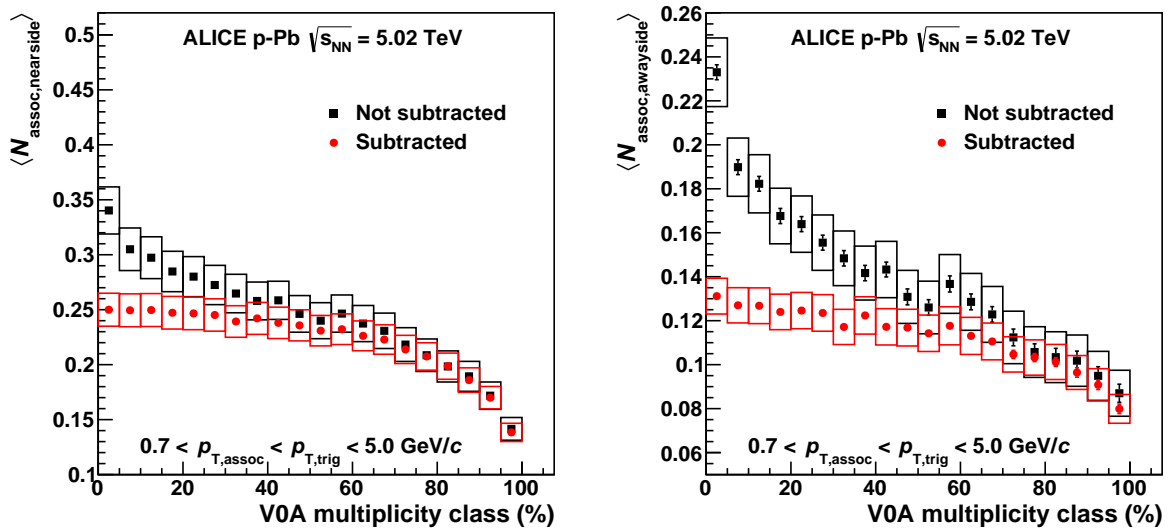}
\caption{\label{fig:double_ridge_subtraction}
Near-side (left panel) and away-side (right panel) per-trigger yields as a function of V0A multiplicity class with (red circles) and without (black squares) subtraction of the long-range correlations. Statistical (lines) and systematic uncertainties (boxes) are shown, even though the statistical ones are mostly smaller than the symbol size.
}
\end{figure}

The effect of the subtraction of the long-range correlations on the measured yields for the V0A estimator is presented in Fig.~\ref{fig:double_ridge_subtraction}, where the near-side and away-side per-trigger yields with (red circles) and without (black squares) long-range correlations subtraction are shown. The yields agree with each other in the multiplicity classes from 50\% to 100\%, consistent with the observation that no significant long-range structure exists in low-multiplicity classes. For higher-multiplicity classes, a difference is observed: the near-side yield increases up to about 0.34 without the subtraction compared to about 0.25 with subtraction. On the away side the value is about 0.23 compared to 0.13. Thus, in the highest multiplicity class, the subtraction procedure removes 30--40\% of the measured yields. The same observation is made for the other multiplicity estimators.

The conclusion drawn earlier, that the hard processes and the number of soft particles show the same evolution with multiplicity, is only valid when the long-range correlations structure is subtracted. 
This observation is consistent with a picture where the minijet-associated yields in \pPb\ collisions originate from the  incoherent fragmentation of multiple parton--parton scatterings,
 while the  long-range correlations  appear unrelated to minijet production.

While the yields give information about the particles produced in a single parton--parton scattering, the uncorrelated seeds calculation (\Eq{eq_us})  provides the number of independent sources of particle production. The uncorrelated seeds are proportional to the number of MPIs in PYTHIA.

\begin{figure}[ht!f]
\centering
\includegraphics[width=0.9\textwidth]{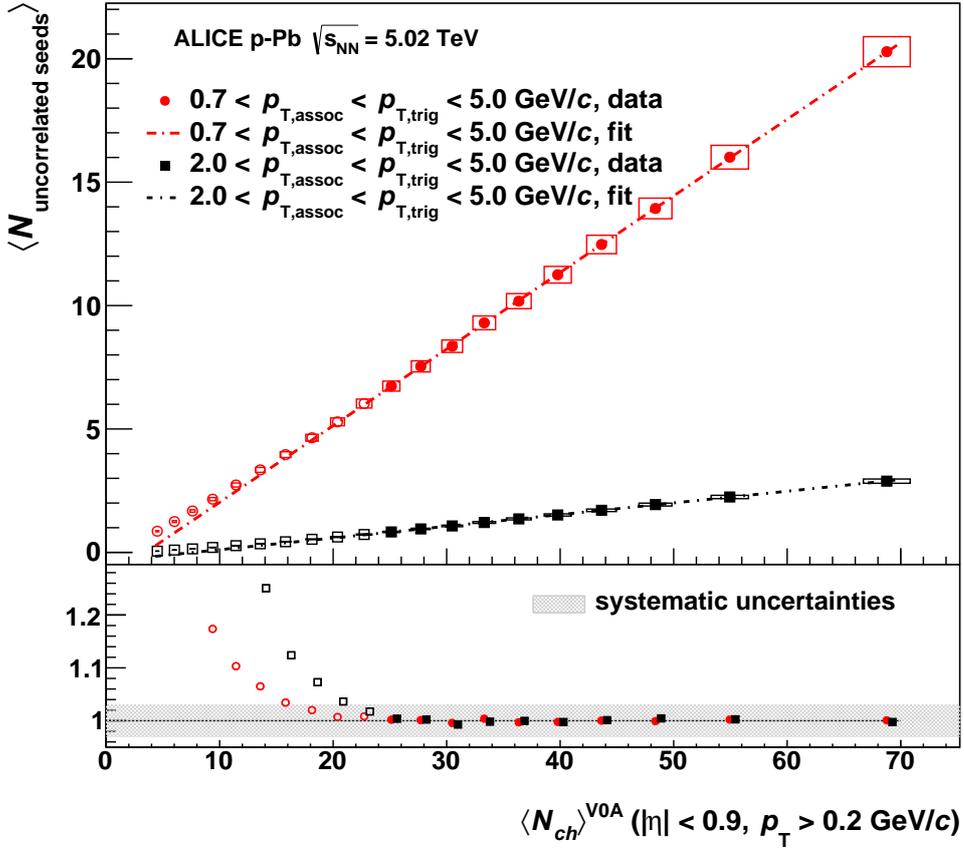}
\caption{\label{fig:uncorrelatedseeds}
Top panel: number of uncorrelated seeds as a function of the midrapidity charged particle multiplicity. Shown are results for two \pt cuts:  \unit[0.7]{\gevc} $< \pta  < \ptt <$ \unit[5.0]{\gevc} (red circles) and \unit[2.0]{\gevc} $< \pta  < \ptt <$ \unit[5.0]{\gevc} (black squares). Each of them is fit with a linear function in the 0--50\% multiplicity classes; open symbols are not included in the fit. Statistical (lines) and systematic uncertainties (boxes) are shown, even though the statistical ones are smaller than the symbol size.
Bottom panel: ratio between the number of uncorrelated seeds and the linear fit functions. Black points are displaced slightly for better visibility.}
\end{figure}

Figure~\ref{fig:uncorrelatedseeds} presents the uncorrelated seeds as a function of the midrapidity charged-particle multiplicity for two \pt cuts. 
In the range \unit[2]{\gevc} $< \pta < \ptt <$ \unit[5]{\gevc}, the number of uncorrelated seeds increases with multiplicity from about 0 to about 3. 
The uncorrelated seeds exhibit a linear increase with midrapidity charged particle multiplicity $\nch$ in particular at high multiplicity. To quantify this behaviour, a linear fit is performed in the 0--50\% multiplicity class and the ratio to the data is presented in the bottom panel.
 
The linear description of the data is valid for $\nch > 20$ while deviations at lower multiplicity are observed.
Deviations from linearity are not surprising as other observables, e.g. the mean $\avg{\pt}$ \cite{Abelev:2013bla} and the $R_{\rm pA}$ \cite{fortheALICE:2013xra}, show a change in dynamics as a function of multiplicity. 
In this $\pt$ range, the uncorrelated seeds are rather similar to the number of particles above a certain $\pt$ threshold as the denominator of \Eq{eq_us} is close to unity.
On the contrary, in the range \unit[0.7]{\gevc} $< \pta < \ptt < $\unit[5.0]{\gevc} the denominator is far from unity.  In this region, the number of uncorrelated seeds increases with multiplicity from about 2 to about 20. The linear description extends over a slightly wider range but a departure is also observed at low multiplicity.

It is interesting to relate the number of uncorrelated seeds to the number of nucleon--nucleon collisions, which in heavy-ion collisions is described successfully by Glauber models~\cite{Glauber} ($N_{\rm coll,\ Glauber}$). However, in \pPb\ collisions, ongoing studies \cite{fortheALICE:2013xra} (to be published in \cite{centralityALICE}) indicate that modifications to the Glauber Monte-Carlo simulations are needed for a correct estimation of the number of hard processes.

\begin{figure}[ht!f]
\centering
\includegraphics[width=0.75\textwidth]{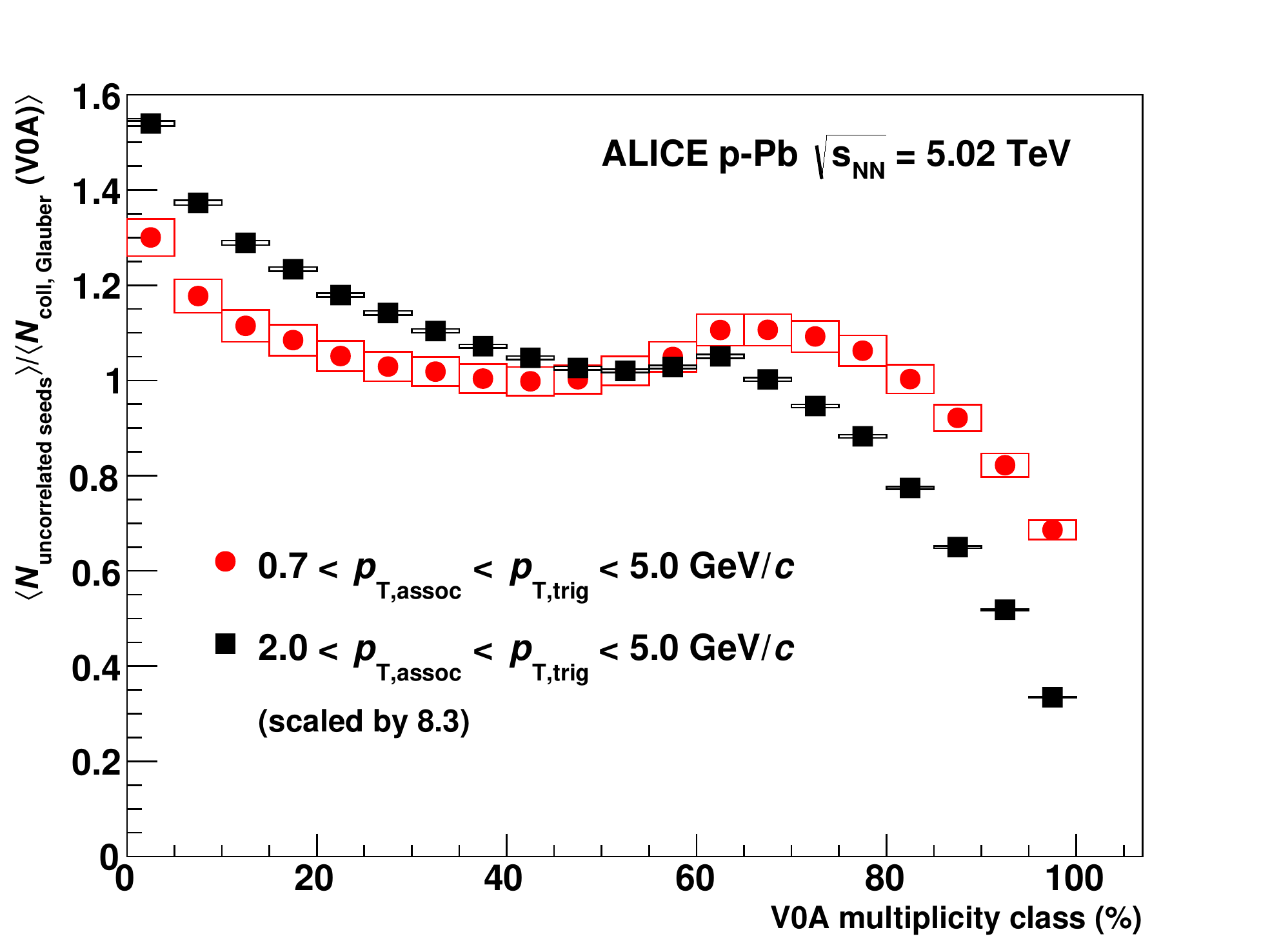}
\caption{\label{fig:usNcoll}
Ratio between uncorrelated seeds and $N_{\rm coll}$ estimated within the Glauber model as a function of V0A multiplicity class. Statistical (lines) and systematic uncertainties (boxes) are shown, even though the statistical ones are smaller than the symbol size. To aid the comparison, the higher \pt range has been scaled by a factor 8.3 to agree with the lower \pt range in the 50--55\% multiplicity class.
}
\end{figure}

Figure~\ref{fig:usNcoll} presents the ratio between uncorrelated seeds and $N_{\rm coll,\ Glauber}$ (calculated with a Glauber Monte-Carlo simulation) as a function of V0A multiplicity class for two \pt cuts. A scaling of the uncorrelated seeds with $N_{\rm coll,\ Glauber}$ within 3\% is observed between 25\% and 55\% multiplicity classes. At higher multiplicity, for the \unit[0.7]{\gevc} $< \pta  < \ptt <$ \unit[5.0]{\gevc} (\unit[2.0]{\gevc} $< \pta  < \ptt <$ \unit[5.0]{\gevc}) range, the ratio between the number of uncorrelated seeds and the number of collisions estimated within the Glauber Monte-Carlo simulations deviates up to 25\% (60\%) from its average. At low multiplicity the deviation is about 30\% (25\%).
This shows that contrary to the expectation for a semi-hard process, the number of uncorrelated seeds is not strictly proportional to the number of binary collisions. For further details, we refer the reader to the publication \Ref{centralityALICE}. Some of these deviations could be due to a bias induced by the centrality estimator.
Monte-Carlo simulations indicate that  by using multiplicity to define event classes, a bias on the mean number of hard collisions per event is introduced: high (low) multiplicity bias towards events with higher (lower) number of semi-hard processes. In addition, low-multiplicity \pPb\ events result from collisions with a larger than average proton--nucleus impact parameter, 
which, for peripheral collisions, corresponds also to a larger than average proton--nucleon impact parameter ~\cite{Jia:2009mq}. 
Therefore, in low-multiplicity collisions the number of MPIs is expected to decrease, which is consistent with the measurement.

\section{Summary}
\label{sec:summary}

Two-particle angular correlations of charged particles have been measured in \pPb\ collisions at $\snn=$~\unit[5.02]{TeV} and expressed as associated yields per trigger particle. Long-range pseudorapidity correlations have been subtracted from the per-trigger yields in order to study the jet-like correlation peaks.
Near-side and away-side jet-like yields are found to be approximately constant over a large range in multiplicity, with the exception of events with low multiplicity. This indicates that at high multiplicity hard processes and number of soft particles have the same evolution with multiplicity.  
These findings are consistent with a picture where independent
parton--parton scatterings with subsequent incoherent fragmentation
produce the measured minijet associated yields, while the ridge yields, which vary with multiplicity, are the result of other sources.
This imposes significant constraints on models which aim at describing \pPb\ collisions. They must reproduce such an incoherent superposition while also describing observations like the ridge structures and the increase of mean \pt with event multiplicity.

The number of uncorrelated seeds increases almost linearly with multiplicity, except at very low multiplicity. Thus, within the measured range, there is no evidence of a saturation in the number of multiple parton interactions.
Furthermore, it is observed that the number of uncorrelated seeds scales only in the intermediate multiplicity region with the number of  binary nucleon--nucleon collisions estimated with Glauber Monte-Carlo simulations, while at high and low multiplicities some biases could possibly cause the scale breaking.

\ifpreprint
\iffull
\newenvironment{acknowledgement}{\relax}{\relax}
\begin{acknowledgement}
\section*{Acknowledgements}
The ALICE Collaboration would like to thank all its engineers and technicians for their invaluable contributions to the construction of the experiment and the CERN accelerator teams for the outstanding performance of the LHC complex.
The ALICE Collaboration gratefully acknowledges the resources and support provided by all Grid centres and the Worldwide LHC Computing Grid (WLCG) collaboration.
The ALICE Collaboration acknowledges the following funding agencies for their support in building and
running the ALICE detector:
State Committee of Science,  World Federation of Scientists (WFS)
and Swiss Fonds Kidagan, Armenia,
Conselho Nacional de Desenvolvimento Cient\'{\i}fico e Tecnol\'{o}gico (CNPq), Financiadora de Estudos e Projetos (FINEP),
Funda\c{c}\~{a}o de Amparo \`{a} Pesquisa do Estado de S\~{a}o Paulo (FAPESP);
National Natural Science Foundation of China (NSFC), the Chinese Ministry of Education (CMOE)
and the Ministry of Science and Technology of China (MSTC);
Ministry of Education and Youth of the Czech Republic;
Danish Natural Science Research Council, the Carlsberg Foundation and the Danish National Research Foundation;
The European Research Council under the European Community's Seventh Framework Programme;
Helsinki Institute of Physics and the Academy of Finland;
French CNRS-IN2P3, the `Region Pays de Loire', `Region Alsace', `Region Auvergne' and CEA, France;
German BMBF and the Helmholtz Association;
General Secretariat for Research and Technology, Ministry of
Development, Greece;
Hungarian OTKA and National Office for Research and Technology (NKTH);
Department of Atomic Energy and Department of Science and Technology of the Government of India;
Istituto Nazionale di Fisica Nucleare (INFN) and Centro Fermi -
Museo Storico della Fisica e Centro Studi e Ricerche "Enrico
Fermi", Italy;
MEXT Grant-in-Aid for Specially Promoted Research, Ja\-pan;
Joint Institute for Nuclear Research, Dubna;
National Research Foundation of Korea (NRF);
CONACYT, DGAPA, M\'{e}xico, ALFA-EC and the EPLANET Program
(European Particle Physics Latin American Network)
Stichting voor Fundamenteel Onderzoek der Materie (FOM) and the Nederlandse Organisatie voor Wetenschappelijk Onderzoek (NWO), Netherlands;
Research Council of Norway (NFR);
Polish Ministry of Science and Higher Education;
National Science Centre, Poland;
 Ministry of National Education/Institute for Atomic Physics and CNCS-UEFISCDI - Romania;
Ministry of Education and Science of Russian Federation, Russian
Academy of Sciences, Russian Federal Agency of Atomic Energy,
Russian Federal Agency for Science and Innovations and The Russian
Foundation for Basic Research;
Ministry of Education of Slovakia;
Department of Science and Technology, South Africa;
CIEMAT, EELA, Ministerio de Econom\'{i}a y Competitividad (MINECO) of Spain, Xunta de Galicia (Conseller\'{\i}a de Educaci\'{o}n),
CEA\-DEN, Cubaenerg\'{\i}a, Cuba, and IAEA (International Atomic Energy Agency);
Swedish Research Council (VR) and Knut $\&$ Alice Wallenberg
Foundation (KAW);
Ukraine Ministry of Education and Science;
United Kingdom Science and Technology Facilities Council (STFC);
The United States Department of Energy, the United States National
Science Foundation, the State of Texas, and the State of Ohio.
\end{acknowledgement}
\ifbibtex
\bibliographystyle{utphys}
\bibliography{biblio}{}
\else
\providecommand{\href}[2]{#2}\begingroup\raggedright\endgroup

\fi
\newpage
\appendix
\section{The ALICE Collaboration}
\label{app:collab}



\begingroup
\small
\begin{flushleft}
B.~Abelev\Irefn{org69}\And
J.~Adam\Irefn{org37}\And
D.~Adamov\'{a}\Irefn{org77}\And
M.M.~Aggarwal\Irefn{org81}\And
M.~Agnello\Irefn{org105}\textsuperscript{,}\Irefn{org88}\And
A.~Agostinelli\Irefn{org26}\And
N.~Agrawal\Irefn{org44}\And
Z.~Ahammed\Irefn{org124}\And
N.~Ahmad\Irefn{org18}\And
I.~Ahmed\Irefn{org15}
% ...
\renewcommand\labelenumi{\textsuperscript{\theenumi}~}

\section*{Affiliation notes}
\renewcommand\theenumi{\roman{enumi}}
\begin{Authlist}
\item \Adef{0}Deceased
\item \Adef{idp1104720}{Also at: St. Petersburg State Polytechnical University}
\item \Adef{idp3001664}{Also at: Department of Applied Physics, Aligarh Muslim University, Aligarh, India}
\item \Adef{idp3684512}{Also at: M.V. Lomonosov Moscow State University, D.V. Skobeltsyn Institute of Nuclear Physics, Moscow, Russia}
\item \Adef{idp3928112}{Also at: University of Belgrade, Faculty of Physics and "Vin\v{c}a" Institute of Nuclear Sciences, Belgrade, Serbia}
\item \Adef{idp4233664}{Permanent Address: Permanent Address: Konkuk University, Seoul, Korea}
\item \Adef{idp4774000}{Also at: Institute of Theoretical Physics, University of Wroclaw, Wroclaw, Poland}
\item \Adef{idp5699264}{Also at: University of Kansas, Lawrence, KS, United States}
\end{Authlist}

\section*{Collaboration Institutes}
\renewcommand\theenumi{\arabic{enumi}~}
\begin{Authlist}

\item \Idef{org1}A.I. Alikhanyan National Science Laboratory (Yerevan Physics Institute) Foundation, Yerevan, Armenia
\item \Idef{org2}Benem\'{e}rita Universidad Aut\'{o}noma de Puebla, Puebla, Mexico
\item \Idef{org3}Bogolyubov Institute for Theoretical Physics, Kiev, Ukraine
\item \Idef{org4}Bose Institute, Department of Physics and Centre for Astroparticle Physics and Space Science (CAPSS), Kolkata, India
\item \Idef{org5}Budker Institute for Nuclear Physics, Novosibirsk, Russia
\item \Idef{org6}California Polytechnic State University, San Luis Obispo, CA, United States
\item \Idef{org7}Central China Normal University, Wuhan, China
\item \Idef{org8}Centre de Calcul de l'IN2P3, Villeurbanne, France
\item \Idef{org9}Centro de Aplicaciones Tecnol\'{o}gicas y Desarrollo Nuclear (CEADEN), Havana, Cuba
\item \Idef{org10}Centro de Investigaciones Energ\'{e}ticas Medioambientales y Tecnol\'{o}gicas (CIEMAT), Madrid, Spain
\item \Idef{org11}Centro de Investigaci\'{o}n y de Estudios Avanzados (CINVESTAV), Mexico City and M\'{e}rida, Mexico
\item \Idef{org12}Centro Fermi - Museo Storico della Fisica e Centro Studi e Ricerche ``Enrico Fermi'', Rome, Italy
\item \Idef{org13}Chicago State University, Chicago, USA
\item \Idef{org14}Commissariat \`{a} l'Energie Atomique, IRFU, Saclay, France
\item \Idef{org15}COMSATS Institute of Information Technology (CIIT), Islamabad, Pakistan
\item \Idef{org16}Departamento de F\'{\i}sica de Part\'{\i}culas and IGFAE, Universidad de Santiago de Compostela, Santiago de Compostela, Spain
\item \Idef{org17}Department of Physics and Technology, University of Bergen, Bergen, Norway
\item \Idef{org18}Department of Physics, Aligarh Muslim University, Aligarh, India
\item \Idef{org19}Department of Physics, Ohio State University, Columbus, OH, United States
\item \Idef{org20}Department of Physics, Sejong University, Seoul, South Korea
\item \Idef{org21}Department of Physics, University of Oslo, Oslo, Norway
\item \Idef{org22}Dipartimento di Fisica dell'Universit\`{a} 'La Sapienza' and Sezione INFN Rome, Italy
\item \Idef{org23}Dipartimento di Fisica dell'Universit\`{a} and Sezione INFN, Cagliari, Italy
\item \Idef{org24}Dipartimento di Fisica dell'Universit\`{a} and Sezione INFN, Trieste, Italy
\item \Idef{org25}Dipartimento di Fisica dell'Universit\`{a} and Sezione INFN, Turin, Italy
\item \Idef{org26}Dipartimento di Fisica e Astronomia dell'Universit\`{a} and Sezione INFN, Bologna, Italy
\item \Idef{org27}Dipartimento di Fisica e Astronomia dell'Universit\`{a} and Sezione INFN, Catania, Italy
\item \Idef{org28}Dipartimento di Fisica e Astronomia dell'Universit\`{a} and Sezione INFN, Padova, Italy
\item \Idef{org29}Dipartimento di Fisica `E.R.~Caianiello' dell'Universit\`{a} and Gruppo Collegato INFN, Salerno, Italy
\item \Idef{org30}Dipartimento di Scienze e Innovazione Tecnologica dell'Universit\`{a} del  Piemonte Orientale and Gruppo Collegato INFN, Alessandria, Italy
\item \Idef{org31}Dipartimento Interateneo di Fisica `M.~Merlin' and Sezione INFN, Bari, Italy
\item \Idef{org32}Division of Experimental High Energy Physics, University of Lund, Lund, Sweden
\item \Idef{org33}Eberhard Karls Universit\"{a}t T\"{u}bingen, T\"{u}bingen, Germany
\item \Idef{org34}European Organization for Nuclear Research (CERN), Geneva, Switzerland
\item \Idef{org35}Faculty of Engineering, Bergen University College, Bergen, Norway
\item \Idef{org36}Faculty of Mathematics, Physics and Informatics, Comenius University, Bratislava, Slovakia
\item \Idef{org37}Faculty of Nuclear Sciences and Physical Engineering, Czech Technical University in Prague, Prague, Czech Republic
\item \Idef{org38}Faculty of Science, P.J.~\v{S}af\'{a}rik University, Ko\v{s}ice, Slovakia
\item \Idef{org39}Frankfurt Institute for Advanced Studies, Johann Wolfgang Goethe-Universit\"{a}t Frankfurt, Frankfurt, Germany
\item \Idef{org40}Gangneung-Wonju National University, Gangneung, South Korea
\item \Idef{org41}Gauhati University, Department of Physics, Guwahati, India
\item \Idef{org42}Helsinki Institute of Physics (HIP), Helsinki, Finland
\item \Idef{org43}Hiroshima University, Hiroshima, Japan
\item \Idef{org44}Indian Institute of Technology Bombay (IIT), Mumbai, India
\item \Idef{org45}Indian Institute of Technology Indore, Indore (IITI), India
\item \Idef{org46}Institut de Physique Nucl\'eaire d'Orsay (IPNO), Universit\'e Paris-Sud, CNRS-IN2P3, Orsay, France
\item \Idef{org47}Institut f\"{u}r Informatik, Johann Wolfgang Goethe-Universit\"{a}t Frankfurt, Frankfurt, Germany
\item \Idef{org48}Institut f\"{u}r Kernphysik, Johann Wolfgang Goethe-Universit\"{a}t Frankfurt, Frankfurt, Germany
\item \Idef{org49}Institut f\"{u}r Kernphysik, Westf\"{a}lische Wilhelms-Universit\"{a}t M\"{u}nster, M\"{u}nster, Germany
\item \Idef{org50}Institut Pluridisciplinaire Hubert Curien (IPHC), Universit\'{e} de Strasbourg, CNRS-IN2P3, Strasbourg, France
\item \Idef{org51}Institute for Nuclear Research, Academy of Sciences, Moscow, Russia
\item \Idef{org52}Institute for Subatomic Physics of Utrecht University, Utrecht, Netherlands
\item \Idef{org53}Institute for Theoretical and Experimental Physics, Moscow, Russia
\item \Idef{org54}Institute of Experimental Physics, Slovak Academy of Sciences, Ko\v{s}ice, Slovakia
\item \Idef{org55}Institute of Physics, Academy of Sciences of the Czech Republic, Prague, Czech Republic
\item \Idef{org56}Institute of Physics, Bhubaneswar, India
\item \Idef{org57}Institute of Space Science (ISS), Bucharest, Romania
\item \Idef{org58}Instituto de Ciencias Nucleares, Universidad Nacional Aut\'{o}noma de M\'{e}xico, Mexico City, Mexico
\item \Idef{org59}Instituto de F\'{\i}sica, Universidad Nacional Aut\'{o}noma de M\'{e}xico, Mexico City, Mexico
\item \Idef{org60}iThemba LABS, National Research Foundation, Somerset West, South Africa
\item \Idef{org61}Joint Institute for Nuclear Research (JINR), Dubna, Russia
\item \Idef{org62}Korea Institute of Science and Technology Information, Daejeon, South Korea
\item \Idef{org63}KTO Karatay University, Konya, Turkey
\item \Idef{org64}Laboratoire de Physique Corpusculaire (LPC), Clermont Universit\'{e}, Universit\'{e} Blaise Pascal, CNRS--IN2P3, Clermont-Ferrand, France
\item \Idef{org65}Laboratoire de Physique Subatomique et de Cosmologie, Universit\'{e} Grenoble-Alpes, CNRS-IN2P3, Grenoble, France
\item \Idef{org66}Laboratori Nazionali di Frascati, INFN, Frascati, Italy
\item \Idef{org67}Laboratori Nazionali di Legnaro, INFN, Legnaro, Italy
\item \Idef{org68}Lawrence Berkeley National Laboratory, Berkeley, CA, United States
\item \Idef{org69}Lawrence Livermore National Laboratory, Livermore, CA, United States
\item \Idef{org70}Moscow Engineering Physics Institute, Moscow, Russia
\item \Idef{org71}National Centre for Nuclear Studies, Warsaw, Poland
\item \Idef{org72}National Institute for Physics and Nuclear Engineering, Bucharest, Romania
\item \Idef{org73}National Institute of Science Education and Research, Bhubaneswar, India
\item \Idef{org74}Niels Bohr Institute, University of Copenhagen, Copenhagen, Denmark
\item \Idef{org75}Nikhef, National Institute for Subatomic Physics, Amsterdam, Netherlands
\item \Idef{org76}Nuclear Physics Group, STFC Daresbury Laboratory, Daresbury, United Kingdom
\item \Idef{org77}Nuclear Physics Institute, Academy of Sciences of the Czech Republic, \v{R}e\v{z} u Prahy, Czech Republic
\item \Idef{org78}Oak Ridge National Laboratory, Oak Ridge, TN, United States
\item \Idef{org79}Petersburg Nuclear Physics Institute, Gatchina, Russia
\item \Idef{org80}Physics Department, Creighton University, Omaha, NE, United States
\item \Idef{org81}Physics Department, Panjab University, Chandigarh, India
\item \Idef{org82}Physics Department, University of Athens, Athens, Greece
\item \Idef{org83}Physics Department, University of Cape Town, Cape Town, South Africa
\item \Idef{org84}Physics Department, University of Jammu, Jammu, India
\item \Idef{org85}Physics Department, University of Rajasthan, Jaipur, India
\item \Idef{org86}Physik Department, Technische Universit\"{a}t M\"{u}nchen, Munich, Germany
\item \Idef{org87}Physikalisches Institut, Ruprecht-Karls-Universit\"{a}t Heidelberg, Heidelberg, Germany
\item \Idef{org88}Politecnico di Torino, Turin, Italy
\item \Idef{org89}Purdue University, West Lafayette, IN, United States
\item \Idef{org90}Pusan National University, Pusan, South Korea
\item \Idef{org91}Research Division and ExtreMe Matter Institute EMMI, GSI Helmholtzzentrum f\"ur Schwerionenforschung, Darmstadt, Germany
\item \Idef{org92}Rudjer Bo\v{s}kovi\'{c} Institute, Zagreb, Croatia
\item \Idef{org93}Russian Federal Nuclear Center (VNIIEF), Sarov, Russia
\item \Idef{org94}Russian Research Centre Kurchatov Institute, Moscow, Russia
\item \Idef{org95}Saha Institute of Nuclear Physics, Kolkata, India
\item \Idef{org96}School of Physics and Astronomy, University of Birmingham, Birmingham, United Kingdom
\item \Idef{org97}Secci\'{o}n F\'{\i}sica, Departamento de Ciencias, Pontificia Universidad Cat\'{o}lica del Per\'{u}, Lima, Peru
\item \Idef{org98}Sezione INFN, Bari, Italy
\item \Idef{org99}Sezione INFN, Bologna, Italy
\item \Idef{org100}Sezione INFN, Cagliari, Italy
\item \Idef{org101}Sezione INFN, Catania, Italy
\item \Idef{org102}Sezione INFN, Padova, Italy
\item \Idef{org103}Sezione INFN, Rome, Italy
\item \Idef{org104}Sezione INFN, Trieste, Italy
\item \Idef{org105}Sezione INFN, Turin, Italy
\item \Idef{org106}SSC IHEP of NRC Kurchatov institute, Protvino, Russia
\item \Idef{org107}SUBATECH, Ecole des Mines de Nantes, Universit\'{e} de Nantes, CNRS-IN2P3, Nantes, France
\item \Idef{org108}Suranaree University of Technology, Nakhon Ratchasima, Thailand
\item \Idef{org109}Technical University of Split FESB, Split, Croatia
\item \Idef{org110}The Henryk Niewodniczanski Institute of Nuclear Physics, Polish Academy of Sciences, Cracow, Poland
\item \Idef{org111}The University of Texas at Austin, Physics Department, Austin, TX, USA
\item \Idef{org112}Universidad Aut\'{o}noma de Sinaloa, Culiac\'{a}n, Mexico
\item \Idef{org113}Universidade de S\~{a}o Paulo (USP), S\~{a}o Paulo, Brazil
\item \Idef{org114}Universidade Estadual de Campinas (UNICAMP), Campinas, Brazil
\item \Idef{org115}University of Houston, Houston, TX, United States
\item \Idef{org116}University of Jyv\"{a}skyl\"{a}, Jyv\"{a}skyl\"{a}, Finland
\item \Idef{org117}University of Liverpool, Liverpool, United Kingdom
\item \Idef{org118}University of Tennessee, Knoxville, TN, United States
\item \Idef{org119}University of Tokyo, Tokyo, Japan
\item \Idef{org120}University of Tsukuba, Tsukuba, Japan
\item \Idef{org121}University of Zagreb, Zagreb, Croatia
\item \Idef{org122}Universit\'{e} de Lyon, Universit\'{e} Lyon 1, CNRS/IN2P3, IPN-Lyon, Villeurbanne, France
\item \Idef{org123}V.~Fock Institute for Physics, St. Petersburg State University, St. Petersburg, Russia
\item \Idef{org124}Variable Energy Cyclotron Centre, Kolkata, India
\item \Idef{org125}Vestfold University College, Tonsberg, Norway
\item \Idef{org126}Warsaw University of Technology, Warsaw, Poland
\item \Idef{org127}Wayne State University, Detroit, MI, United States
\item \Idef{org128}Wigner Research Centre for Physics, Hungarian Academy of Sciences, Budapest, Hungary
\item \Idef{org129}Yale University, New Haven, CT, United States
\item \Idef{org130}Yonsei University, Seoul, South Korea
\item \Idef{org131}Zentrum f\"{u}r Technologietransfer und Telekommunikation (ZTT), Fachhochschule Worms, Worms, Germany
\end{Authlist}
\endgroup

\else
\ifbibtex
\bibliographystyle{utphys}
\bibliography{biblio}{}
\else

\fi
\fi
\else
\iffull
\vspace{0.5cm}

\ifbibtex
\bibliographystyle{model1-num-names}
\bibliography{biblio}{}
\else
\input{refpaper.tex}
\fi
\else
\ifbibtex
\bibliographystyle{model1-num-names}
\bibliography{biblio}{}
\else
\input{refpaper.tex}
\fi
\fi
\fi
\end{document}